\documentclass[onecolumn,amsmath,amssymb,nofootinbib,pra]{revtex4} 

\usepackage{amsmath}
\usepackage{graphicx}
\usepackage{graphics}
\usepackage{dcolumn}
\usepackage{bm}
\usepackage{hyperref}
\usepackage{xcolor}

\def\BEq{\begin{equation}}
\def\EEq{\end{equation}}
\def\BEqA{\begin{eqnarray}}
\def\EEqA{\end{eqnarray}}
\def\BEn{\begin{enumerate}}
\def\EEn{\end{enumerate}}
\def\BWT{\begin{widetext}}
\def\EWT{\end{widetext}}

\begin{document}

\title{Strongly anisotropic Rytova-Keldysh interaction and the ground 
state of 2D excitons}

\author{Andrei Galiautdinov}
\email{ag1@uga.edu}
\affiliation{
Department of Physics and Astronomy, 
University of Georgia, Athens, Georgia 30602, USA
}

\date{\today}

\begin{abstract}
The classic Rytova-Keldysh potential describes the non-local 
dielectric screening of Coulomb interactions in ultrathin 
two-dimensional (2D) materials. Recently, the corresponding 
potential for arbitrary in-plane anisotropy was derived in 
integral form, with numerical studies suggesting that an effective 
isotropic approximation remains robust for highly directional 
systems. In this paper we provide the analytical foundation 
for these observations by mapping the complete spatial 
landscape of the strongly anisotropic Rytova-Keldysh interaction. 
By applying the method of steepest descent and momentum-space 
coordinate scaling to the exact one-dimensional integral representation, 
we derive closed-form asymptotic expressions for the potential 
across all spatial regimes. We find that the intermediate-range 
screening exhibits a non-trivial amplitude scaling driven by 
the direction of weakest polarizability, while the short-range 
limit produces an anisotropic logarithmic well governed by 
the geometric mean of the principal polarizabilities. Finally, 
utilizing this short-range confinement, we implement an 
anisotropic Gaussian variational ansatz to solve the Wannier 
equation, providing a closed-form analytical expression for 
the exciton ground-state binding energy that explicitly 
captures the competition between the effective mass and 
dielectric tensors.
\end{abstract}

\maketitle

\section{Introduction}

The theoretical description of excitons and many-body 
correlations in two-dimensional (2D) semiconductors 
relies fundamentally on the Rytova-Keldysh model 
potential \cite{Rytova1967, Keldysh1979, Keldysh1997}, 
\BEq 
V_{\rm RK}(r)
=
\frac{\pi q q'}{2 r_0}
\left[H_0\left(\frac{\kappa r}{r_0}\right)
-
Y_0\left(\frac{\kappa r}{r_0}\right)\right],
\quad
r_0 \equiv \frac{\epsilon d}{2},
\EEq 
which accounts for the pronounced non-local dielectric 
screening inherent to atomically thin geometries. The 
precise formulation of this interaction remains critical 
for understanding interaction-driven phenomena in 2D 
materials, ranging from magnetoexciton hybridization 
\cite{Le2026} and exciton transport \cite{Chang2026, Seksaria2024} 
to DFT+U+V renormalizations of spin-orbit splitting in 
isotropic transition metal dichalcogenides \cite{Rozhansky2026}. 
The recent discovery of 2D materials with extremely low crystal 
symmetry---such as black phosphorus, tellurene, quasi-1D perovskite 
chalcogenides, the heavily anisotropic van der Waals magnet CrSBr, 
as well as others (e.\ g., \cite{Brunetti2019, Ermolaev2021, Hwangbo2021, 
Carre2024, Meineke2024, Engdahl2025, Cui2025, Qian2025, 
Semina2025, Smolenski2025, Rizzo2025, Komar2025, Sears2025, 
Li2025, deOliveira2025, Ermolaev2025, Martins2026, 
Ramasubramaniam2026})---provides a compelling motivation 
for extending this potential to include the in-plane dielectric anisotropy. 

In 2019, generalization of the Rytova-Keldysh interaction 
was derived in the anisotropic case \cite{Galiautdinov2019}. 
By employing a Fourier series expansion, the isotropic Struve 
and Bessel functions were shown to smoothly deform into 
directional multipole moments. In addition, explicit asymptotic 
expressions for the interaction potential were provided 
in closed analytical form in the case of weak anisotropy.  
Subsequently, in Ref.\ \cite{Kamban2020}, that approach 
was extended to arbitrary anisotropy by linearizing 
the effective dielectric function and expressing 
the corresponding potential in a 1D integral form. 
Notably, the authors demonstrated numerically that 
an effective isotropic interaction serves as a robust 
approximation for modeling 2D excitons, suggesting 
that observable anisotropies are heavily influenced by the 
effective mass tensor rather than the dielectric tensor alone. 

While this numerical observation is practically useful, 
an analytical derivation elucidating how the strongly 
anisotropic dielectric tensor reduces to an effective 
isotropic-like potential at relevant length scales has been 
missing. In this paper we provide this mathematical 
foundation. By analyzing the exact arbitrary-anisotropy 
1D integral representation, we extract the asymptotic 
behavior of the potential across all spatial 
regimes. Utilizing the method of steepest descent for 
the intermediate scale and momentum-space coordinate 
scaling for the short-range limit, we demonstrate how 
the effective screening length scales with the principal 
polarizabilities. Finally, we implement the short-range 
potential in an anisotropic variational approach 
(cf., e.\ g., \cite{Gomes2022, delValle2023, Rudenko2024}) 
to the 2D Wannier equation, yielding a closed-form analytical 
expression for the exciton ground-state binding energy.

\section{Exact One-Dimensional Representation}

The electrostatic potential energy of interaction between 
two charges $q$ and $q'$ in an anisotropic 2D material 
of thickness $d$, embedded in an asymmetric environment 
described by the avearge dielectric constant, 
$\kappa = (\epsilon_{\rm top}+\epsilon_{\rm bottom})/2$, 
is given by (see \cite{Galiautdinov2019} for details; 
also \cite{Kamban2020} and \cite{Ceferino2020}),
\begin{equation}
\label{eq:Vxy}
V(x, y) = \frac{qq'}{2\pi} 
\iint 
\frac{e^{i (k_x x + k_y y)}}{\kappa k + r_x k_x^2 + r_y k_y^2} 
dk_x dk_y,
\end{equation}
where $r_x=\epsilon_x d/2$ and $r_y=\epsilon_y d/2$ 
are the effective 2D screening lengths along the principal 
crystallographic axes (characterized by the dielectric 
constants $\epsilon_x$ and $\epsilon_y$ of the material), and 
$k = \sqrt{k_x^2 + k_y^2}$. Instead of expanding the denominator 
(which was done in Refs.\ \cite{Galiautdinov2019, Kamban2020}), we transition 
to polar coordinates in both momentum and real space via
$k_x = k \cos\theta$, $k_y = k \sin\theta$, and
$x = r \cos\phi$, $y = r \sin\phi$, respectively.
This gives,
\begin{equation}
\label{eq:exactVrphi-2d}
V(r, \phi) 
= 
\frac{qq'}{2\pi} \int_0^{2\pi} d\theta 
\int_0^\infty \frac{\cos(k r \cos(\theta - \phi))}{\kappa + k \rho(\theta)} dk,
\end{equation}
where we have defined the angular screening function,
\BEq
\rho(\theta) = r_x \cos^2\theta + r_y \sin^2\theta .
\EEq 
The inner integral over $k$ is a Laplace-type transform. Letting 
\BEq
u(r, \theta, \phi) = r |\cos(\theta - \phi)| \geq 0,
\EEq
and taking into account that 
$\cos(k r \cos(\theta - \phi))=\cos(k r |\cos(\theta - \phi)|)$,
we perform the $k$-integration, 
\begin{align}
\int_0^\infty 
\frac{\cos(k r \cos(\theta - \phi))}{\kappa + k \rho} dk
&=
\int_0^\infty 
\frac{\cos(k r |\cos(\theta - \phi)|)}{\kappa + k \rho} dk
=
\int_{\kappa}^\infty 
\frac{\cos\left(\frac{(\xi-\kappa)u}{\rho}\right)}{\xi } \frac{d\xi}{\rho}
\nonumber \\
&=
\frac{1}{\rho}
\int_{\kappa}^\infty 
\left[
\frac{\cos\left(\frac{\xi u}{\rho}\right)
\cos\left(\frac{\kappa u}{\rho}\right)}{\xi } 
+
\frac{\sin\left(\frac{\xi u}{\rho}\right)
\sin\left(\frac{\kappa u}{\rho}\right)}{\xi} 
\right]d\xi
\nonumber \\
&=
\frac{1}{\rho}
\left[ 
\cos\left(\frac{\kappa u}{\rho}\right)
\int_{\kappa u/\rho}^\infty \frac{\cos t}{t } dt
+
\sin\left(\frac{\kappa u}{\rho}\right)
\int_{\kappa u/\rho}^\infty \frac{\sin t}{t} dt 
\right]
\nonumber \\
&=
\frac{1}{\rho} 
\left[ -\cos\left(\frac{\kappa u}{\rho}\right) \text{Ci}\left(\frac{\kappa u}{\rho}\right) 
- \sin\left(\frac{\kappa u}{\rho}\right) \text{si}\left(\frac{\kappa u}{\rho}\right) \right],
\end{align}
which reduces
the infinite 2D oscillatory integral in Eq.\ (\ref{eq:exactVrphi-2d}) 
to a finite 1D integral form,
\begin{equation}
\label{eq:exactVrphi}
V(r, \phi) = \frac{qq'}{2\pi} \int_0^{2\pi} I(r, \theta, \phi) d\theta,
\quad
I(r, \theta, \phi) = \frac{1}{\rho(\theta)} 
\left[ -\cos(Z) \text{Ci}(Z) - \sin(Z) \text{si}(Z) \right],
\quad
Z \equiv \frac{\kappa u(r, \theta, \phi)}{\rho(\theta)},
\end{equation}
where $\text{Ci}$ and $\text{si}$ are the standard 
cosine and sine integrals,
\BEq
\text{Ci}(x)=-\int_{x}^{\infty}\frac{\cos(t)}{t}dt,
\quad
\text{si}(x)=-\int_{x}^{\infty}\frac{\sin(t)}{t}dt.
\EEq
Within the limitations of the original 2D model, formula (\ref{eq:exactVrphi}) is exact for any arbitrary degree of anisotropy,
which may be contrasted with Eq.\ (A30) found
in Ref.\  \cite{Kamban2020}.

\section{Short-Range Asymptotics and The Anisotropic Logarithmic Well}

To calculate observable quantities such as the exciton ground-state 
binding energy, we must extract the short-range limit of the potential 
at $r \ll r_x, r_y$. Here we demonstrate how that can be done directly 
from the exact one-dimensional integral representation, 
Eq.\ (\ref{eq:exactVrphi}); cf.\ Appendix \ref{sec:altVshort}. 
In this regime, the dimensionless argument 
of the auxiliary function,
$Z = \kappa r |\cos(\theta - \phi)|/\rho(\theta)$,
approaches zero. Substituting the small-argument Taylor expansions 
of the sine and cosine integrals, 
$\text{Ci}(Z) \approx \gamma_{\rm E} + \ln Z$, 
$\text{si}(Z) \approx -\frac{\pi}{2}$,
where $\gamma_{\rm E}$ is the Euler-Mascheroni constant, 
the bracketed term of the integrand simplifies to
$-\cos(Z)\text{Ci}(Z) - \sin(Z)\text{si}(Z) 
\approx 
-\left( \gamma_{\rm E} + \ln Z \right)$,
and the integrand in Eq.\ (\ref{eq:exactVrphi}) becomes
\begin{equation}
I(r, \theta, \phi) 
\approx 
-\frac{1}{\rho(\theta)} 
\left[ \gamma_{\rm E} 
+ \ln\left( \frac{\kappa r |\cos(\theta - \phi)|}{\rho(\theta)} \right) \right].
\end{equation}
Inserting this into the full angular integral, Eq.\ (\ref{eq:exactVrphi}), 
the potential separates into
\begin{equation}
\label{eq:Vshort_split}
V_{\rm short}(r, \phi) 
\approx 
-\frac{qq'}{2\pi} \ln\left( \frac{r}{\sqrt{r_x r_y}} \right) 
\int_0^{2\pi} \frac{d\theta}{\rho(\theta)} 
- \frac{qq'}{2\pi} 
\int_0^{2\pi} \frac{\ln|\cos(\theta - \phi)|}{\rho(\theta)} d\theta 
- \frac{qq'}{2\pi} 
\int_0^{2\pi}
\frac{1}{\rho(\theta)} 
\ln\left( \frac{e^{\gamma_{\rm E}} \kappa \sqrt{r_x r_y}}{\rho(\theta)} \right) d\theta.
\end{equation}

To evaluate these angular integrals analytically, we introduce 
a continuous change of variables to a scaled angle $\theta'$, 
defined by $\tan\theta = \sqrt{r_x/r_y} \tan\theta'$.
Differentiating this relation provides the correspondence 
between the differential elements,
\begin{equation}
d\theta 
= 
\frac{\sqrt{r_x r_y}}{r_y \cos^2\theta' + r_x \sin^2\theta'} d\theta' 
=  
\frac{\rho(\theta)}{\sqrt{r_x r_y}} d\theta',
\end{equation}
where we used the relationship,
\BEq
r_x \cos^2\theta + r_y \sin^2\theta
=
\frac{r_x r_y}{r_y \cos^2\theta' + r_x \sin^2\theta'},
\EEq
which immediately solves the first integral in Eq.\ (\ref{eq:Vshort_split}) as 
\begin{equation}
\int_0^{2\pi} {d\theta}/{\rho(\theta)} 
= 
\int_0^{2\pi} {d\theta'}/{\sqrt{r_x r_y}}
=
{2\pi}/{\sqrt{r_x r_y}}.
\end{equation}

To evaluate the second integral in Eq.\ (\ref{eq:Vshort_split}), we note that 
$\cos\theta = \sqrt{r_y}\cos\theta' / N(\theta')$,
$\sin\theta = \sqrt{r_x}\sin\theta' / N(\theta')$, where
$ N(\theta') \equiv \sqrt{r_y \cos^2\theta' + r_x \sin^2\theta'}$.
Applying the angle addition formula, the logarithmic term transforms to
\begin{align}
\label{eq:ln|cos|}
\ln|\cos(\theta - \phi)| 
&= \ln\left| \frac{\sqrt{r_y}\cos\phi \cos\theta' 
+ \sqrt{r_x}\sin\phi \sin\theta'}{N(\theta')} \right| 
\nonumber \\
&= 
\frac{1}{2}\ln\left(\frac{r_y \cos^2\phi + r_x \sin^2\phi}{\sqrt{r_xr_y}}\right) 
+ \ln|\cos(\theta' - \alpha)| 
- \frac{1}{2}\ln \left(\frac{N^2(\theta')}{\sqrt{r_xr_y}}\right),
\end{align}
where $\alpha$ is a phase angle defined by $\tan \alpha = \sqrt{r_x/r_y}\tan \phi$.
When integrating over $d\theta'$ from $0$ to $2\pi$, the second term in (\ref{eq:ln|cos|})
yields the standard definite integral, $\int_0^{2\pi} \ln|\cos x| dx = -2\pi \ln 2$. 
The third term in (\ref{eq:ln|cos|}), evaluated via the identity 
\BEq
\label{eq:usefuIntegral}
\int_0^{2\pi} \ln(A\cos^2 x + B\sin^2 x) dx = 4\pi \ln[(\sqrt{A}+\sqrt{B})/2], 
\EEq
produces
\begin{equation}
-\frac{1}{2} \int_0^{2\pi} \ln \left(\frac{N^2(\theta')}{\sqrt{r_xr_y}}\right) d\theta' 
= 
-\pi \ln\left[ \frac{(\sqrt{r_x} + \sqrt{r_y})^2}{4\sqrt{r_x r_y}} \right].
\end{equation}
Similarly, to integrate the final term from Eq.\ (\ref{eq:Vshort_split}), we use substitution, $\rho(\theta) = r_x r_y / N^2(\theta')$, which 
generates analogous definite integrals that evaluate to 
$2\pi(\gamma_{\rm E} + \ln \kappa) + 2\pi \ln[(\sqrt{r_x} + \sqrt{r_y})^2 / (4\sqrt{r_x r_y})]$. 

Finally, substituting all these evaluated components back into 
Eq.\ (\ref{eq:Vshort_split}), and transforming back to Cartesian coordinates, we obtain the full analytical form 
of the short-range potential,
\begin{equation}
\label{eq:Vshort_final}
V_{\rm short}(x, y) 
\approx 
V_0
 \ln\left[
\frac{e^{2\gamma_{\rm E}} \kappa^2 (\sqrt{r_x} + \sqrt{r_y})^2}{16 r_x r_y}
\left( \frac{x^2}{r_x} + \frac{y^2}{r_y} \right)
\right],
\quad
V_0 \equiv -\frac{qq' }{2\sqrt{r_x r_y}} > 0.
\end{equation}
This shows that the effective depth of the confining potential 
is governed by the geometric mean of the polarizabilities, 
motivating the use of elliptical equipotential contours that 
define the symmetries for our variational ansatz to be 
introduced in Sec.\ \ref{sec:groundState}.
Notice also that in the isotropic limit characterized by
$r_x=r_y\equiv r_0 \equiv \epsilon d/2$,
we recover the standard short range expression  
(cf.\ Eq.\ (3) in Ref.\ \cite{Keldysh1979}),
\begin{equation}
\label{eq:Vshort_Keldysh}
V_{\rm short}^{\text{(isotropic)}}(r) 
=
\frac{2qq'}{\epsilon d}
\left[
 \ln\left(\frac{\epsilon d}{\kappa r}\right)-\gamma_{\rm E}
\right].
\end{equation}
As was pointed by Keldysh in his original 1979 paper
\cite{Keldysh1979}, this specific logarithmic behavior 
occurs when the system parameters satisfy the strict
inequality,
$\left({\kappa}/{\epsilon}\right)^2 \ll {d}/{a_{\rm B}}\ll 1$,
where $a_{\rm B} \equiv {\epsilon \hbar^2}/{(m(-qq'))}$
is the effective Bohr radius of the bulk exciton.

\section{Method of Steepest Descent and Intermediate Scaling}

The true analytical challenge lies in the intermediate spatial regime, 
$r \sim r_x$, under strong anisotropy, $r_x \gg r_y$, 
while assuming that $\kappa r/r_y$ is finite. 
In this limit, the angular screening function $\rho(\theta)$ is highly 
elongated, causing the inverse angular screening $1/\rho(\theta)$ to 
exhibit sharp maxima along the $y$-axis, at $\theta_0 = \pi/2$ and 
$3\pi/2$. We evaluate the exact angular integral using Laplace's 
method of steepest descent. We expand 
$\rho(\theta)$ around the primary peak $\theta_0 = \pi/2$ by 
defining 
\BEq
\theta = \pi/2 + \delta.
\EEq 
To second order in $\delta$, 
the screening function is
\BEq
\rho(\delta) \approx r_y + (r_x - r_y)\delta^2.
\EEq
For $r_x \gg r_y$, the integrand takes the form of a sharp 
Lorentzian-like peak. Approximating the smoothly varying 
functions in $I(r,\theta,\phi)$ by their peak values at $\theta_0$, 
the full integral is dominated by two identical 
localized contributions (hence the extra factor of 2),
providing the intermediate-range asymptotic potential,
\begin{equation}
\label{eq:Vintermediate}
V_{\rm intermediate}(r, \phi) 
\approx 
\frac{qq'}{\pi} {\cal I}(r,\pi/2,\phi) 
\int_{-\infty}^{\infty} \frac{d\delta}{r_y + (r_x - r_y)\delta^2} 
=
\frac{qq' {\cal I}(r,\pi/2, \phi)}{\sqrt{r_y (r_x - r_y)}},
\end{equation}
where
\BEq
\label{eq:Ipiover2}
{\cal I}(r, \pi/2, \phi) 
= 
-\cos\left(\frac{\kappa r|\sin \phi|}{r_y}\right) 
\text{Ci}\left(\frac{\kappa r|\sin \phi|}{r_y}\right) 
- \sin\left(\frac{\kappa r|\sin \phi|}{r_y}\right) 
\text{si}\left(\frac{\kappa r|\sin \phi|}{r_y}\right).
\EEq
This asymptotic expression is highly 
accurate for the majority of the 2D plane, but breaks down in 
a narrow angular wedge along the $x$-axis, as $\phi \to 0, \pi$. 
At these specific angles, the logarithmic singularity inherent 
to the cosine integral coincides with the saddle point at 
$\theta_0=\pi/2$, violating the Laplace method's assumption 
of a slowly varying prefactor ${\cal I}$.
Eq.\ (\ref{eq:Vintermediate}) establishes a non-trivial physical result: 
in strongly anisotropic 2D materials, the effective intermediate 
screening does not scale with the arithmetic or geometric mean 
of the polarizabilities, but rather scales with $\sqrt{r_y (r_x - r_y)}$. 
The direction of the weakest polarizability controls the amplitude 
of the potential.

\section{Long-Range Asymptotics}

For completeness, we must also verify the long-range asymptotic 
behavior of the potential at $r \gg r_x, r_y$. To extract this limit 
from the exact one-dimensional representation, we first define 
the auxiliary function,
\BEq
{\cal I}(Z) \equiv -\cos(Z) \text{Ci}(Z) - \sin(Z) \text{si}(Z).
\EEq 
Applying the standard large-argument expansions for the cosine and sine integrals, 
$\text{Ci}(Z) \approx {\sin Z}/{Z}- {\cos Z}/{Z^2}$,
$\text{si}(Z) \approx -{\cos Z}/{Z} - {\sin Z}/{Z^2}$,
we find that the leading $1/Z$ terms cancel exactly, yielding 
${\cal I}(Z) \approx 1/Z^2$ for $Z \gg 1$. Substituting this directly 
into the angular integral (\ref{eq:exactVrphi}) results in an 
integrand proportional to $1/(r^2 \cos^2(\theta - \phi))$, 
which introduces non-integrable singularities at 
$\theta = \phi \pm \pi/2$ (in addition to the physically 
meaningless $1/r^2$ behavior at large distances). 
This divergence indicates that the limit $r \to \infty$ and the integration 
over $\theta$ do not commute. Because 
$Z = \kappa r |\cos(\theta - \phi)| / \rho(\theta)$, the argument 
$Z$ vanishes at the poles $\theta_0 = \phi \pm \pi/2$ 
regardless of the magnitude of $r$, and the dominant contribution 
to the integral at large $r$ arises from the infinitesimally 
narrow angular regions surrounding these poles, where the large-$Z$ 
expansion is invalid.

To extract the correct long-range behavior, we employ 
a steepest-descent evaluation around the poles
(see Appendix \ref{sec:altVlong} for alternative derivation). 
We introduce 
a small angular deviation $\delta$ such that 
\BEq
\theta = \phi \pm \pi/2 + \delta,
\EEq 
so that in the immediate vicinity of 
either pole, we can approximate 
\BEq
|\cos(\theta - \phi)| \approx |\delta|.
\EEq
The angular screening function can be treated as a constant 
evaluated at the poles,
\begin{equation}
\rho_\perp \equiv \rho(\phi \pm \pi/2) = r_x \sin^2\phi + r_y \cos^2\phi.
\end{equation}
Because the dominant contributions are tightly localized, we can 
extend the integration limits for $\delta$ to infinity. Summing 
the symmetric contributions from both poles, $\phi + \pi/2$ 
and $\phi - \pi/2$, the asymptotic potential becomes
\begin{equation}
V_{\rm long}(r, \phi) 
\approx 
\frac{qq'}{\pi} \int_{-\infty}^{\infty} 
\frac{d\delta}{\rho_\perp} {\cal I}\left( \frac{\kappa r |\delta|}{\rho_\perp} \right).
\end{equation}
Taking advantage of the even parity of the integrand, we restrict 
the domain to $\delta \ge 0$ and introduce the change of 
variables, 
\BEq
z = \kappa r \delta / \rho_\perp.
\EEq
The differential element transforms as 
$d\delta = (\rho_\perp / \kappa r) dz$, 
which cancels the dimensional $\rho_\perp$ parameter in the denominator, 
giving
\BEq
V_{\rm long}(r, \phi) 
\approx 
\frac{2qq'}{\pi \kappa r} \int_0^\infty {\cal I}(z) dz.
\EEq
Using the Laplace integral 
representation of our auxiliary function, 
\BEq
{\cal I}(z) = \int_0^\infty \frac{t e^{-zt}}{1+t^2} dt,
\EEq
and taking into account that 
\BEq
\int_0^\infty {\cal I}(z) dz 
= \int_0^\infty \left[ \int_0^\infty \frac{t e^{-zt}}{1+t^2} dt \right] dz
= \int_0^\infty \frac{t}{1+t^2} \left[ \int_0^\infty e^{-zt} dz \right] dt 
= \int_0^\infty \frac{t}{1+t^2} \left( \frac{1}{t} \right) dt 
= \int_0^\infty \frac{1}{1+t^2} dt = \frac{\pi}{2}.
\EEq
we get
\begin{equation}
V_{\rm long}(r, \phi) \approx \frac{qq'}{\kappa r},
\end{equation}
correctly recovering the isotropic 3D Coulomb limit.
Physically, this confirms that the intrinsic two-dimensional anisotropy 
governing the short-range exciton confinement is entirely 
screened out in the macroscopic limit.

\section{Variational Ground State of the Wannier Equation}
\label{sec:groundState}

To calculate the excitonic ground state, we turn to the short-range 
potential, $V_{\rm short}(x, y)$, given in (\ref{eq:Vshort_final}). 
This deep logarithmic well binds the electron-hole pair. 
The corresponding effective-mass Wannier Hamiltonian is given by
\begin{equation}
H 
= 
-\frac{\hbar^2}{2m_x}\frac{\partial^2}{\partial x^2} 
- \frac{\hbar^2}{2m_y}\frac{\partial^2}{\partial y^2} 
+ V_{\rm short}(x, y),
\end{equation}
where $m_x$ and $m_y$ are the effective masses along 
the $x$ and $y$ directions, respectively. Because the potential 
exhibits elliptical symmetry, the standard hydrogenic $1s$ 
exponential is inadequate. Therefore, we construct an anisotropic 
2D Gaussian trial wavefunction,
\begin{equation}
\Psi(x, y) 
= 
\frac{1}{\sqrt{\pi a b}} 
\exp\left( -\frac{x^2}{2a^2} - \frac{y^2}{2b^2} \right),
\end{equation}
and enforce the variational parameters $a$ and $b$ to scale 
geometrically with the lattice anisotropy by defining 
$a = \lambda r_x$, $b = \lambda r_y$,
where $\lambda$ is a dimensionless scaling parameter. 
The expectation value of the kinetic energy is then immediately
found to be
\begin{equation}
\langle T \rangle 
= 
\frac{\hbar^2}{4\lambda^2} 
\left( \frac{1}{m_x r_x^2} + \frac{1}{m_y r_y^2} \right) 
\equiv \frac{T_0}{\lambda^2}.
\end{equation}
The expectation value of the logarithmic potential is given by
$\langle V \rangle = \iint |\Psi(x,y)|^2 V_{\rm short}(x,y) dx dy$.
To perform this integration, we transform to scaled polar 
coordinates defined by 
$x = \lambda r_x \rho \cos\varphi$,
$y = \lambda r_y \rho \sin\varphi$, resulting in
\begin{equation}
\langle V \rangle 
= 
\frac{V_0}{\pi} 
\int_0^{2\pi} d\varphi \int_0^\infty \rho d\rho e^{-\rho^2} 
\ln\left[ 
\frac{e^{2\gamma_{\rm E}} \kappa^2 (\sqrt{r_x} + \sqrt{r_y})^2}{16 r_x r_y} 
\lambda^2 \rho^2 \left( r_x \cos^2\varphi + r_y \sin^2\varphi \right) 
\right].
\end{equation}
Decomposing the logarithm isolates the integration 
involving the scaling parameter $\lambda$, yielding 
$V_0 \ln(\lambda^2)$. The remaining terms constitute 
a geometry-dependent constant $C$, given by
\begin{align}
C 
&= 
\ln\left[ \frac{e^{2\gamma_{\rm E}} \kappa^2 (\sqrt{r_x} + \sqrt{r_y})^2}{16 r_x r_y} \right] 
+ \frac{1}{\pi} \int_0^{2\pi} d\varphi \int_0^\infty \rho d\rho e^{-\rho^2} \ln(\rho^2) 
+ \frac{1}{\pi} \int_0^\infty \rho d\rho e^{-\rho^2} \int_0^{2\pi} 
d\varphi \ln\left( r_x \cos^2\varphi + r_y \sin^2\varphi \right).
\end{align}
Substituting $t = \rho^2$, the radial integral evaluates 
to $\int_0^\infty e^{-t} \ln(t) dt = -\gamma_{\rm E}$. 
Applying the identity given in Eq.\ (\ref{eq:usefuIntegral}) 
to the angular integral yields $\ln[(\sqrt{r_x} + \sqrt{r_y})^2 / 4]$. 
Summing these contributions, we find,
\begin{equation}
C = 2\gamma_{\rm E} - \gamma_{\rm E} 
+ \ln\left[ \frac{\kappa^2 (\sqrt{r_x} + \sqrt{r_y})^2}{16 r_x r_y} \right] 
+ \ln\left[ \frac{(\sqrt{r_x} + \sqrt{r_y})^2}{4} \right] 
= \gamma_{\rm E} + \ln\left[ \frac{\kappa^2 (\sqrt{r_x} + \sqrt{r_y})^4}{64 r_x r_y} \right].
\end{equation}
Thus, the expectation value of the potential reduces to
\begin{equation}
\langle V \rangle = V_0 \ln(\lambda^2) + V_0 C.
\end{equation}

The total energy functional 
$E(\lambda) = \langle T \rangle + \langle V \rangle$ 
is next minimized by taking $\partial E / \partial \lambda = 0$,
\begin{equation}
-\frac{2 T_0}{\lambda^3} + \frac{2 V_0}{\lambda} 
= 
0 \implies \lambda_{\min}^2 
= 
\frac{T_0}{V_0}.
\end{equation}
Substituting the equilibrium scaling factor back into 
the energy functional leads to
\BEq
E_{\rm gs} = T_0 (V_0/T_0) + V_0 \ln(T_0/V_0) + V_0 C,
\EEq 
which results in the closed-form exciton ground-state energy,
\begin{equation}
\label{eq:Egs}
E_{\rm gs} 
= 
V_0 \left[ 
1 + \gamma_{\rm E} 
+ \ln\left( \frac{\kappa^2 (\sqrt{r_x} + \sqrt{r_y})^4}{64 r_x r_y} \right) 
+ \ln\left( \frac{\hbar^2}{4 V_0} 
\left( \frac{1}{m_x r_x^2} + \frac{1}{m_y r_y^2} \right) \right) 
\right].
\end{equation}
In the isotropic limit 
($r_x=r_y\equiv r_0 \equiv \epsilon d/2$,
$m_x=m_y\equiv m$) this becomes
\begin{equation}
\label{eq:Egs_Keldysh}
E_{\rm gs}^{\text{(isotropic)}} 
=
\frac{qq'}{\epsilon d}
\left[
 \ln\left(\frac{\epsilon^2 d}{\kappa^2 a_{\rm B}}\right) 
 + \ln 2 - \gamma_{\rm E} - 1
\right].
\end{equation}

Eq. (\ref{eq:Egs}) gives a closed-form analytical expression 
for and provides a reasonably good approximation to the ground 
state energy in the isotropic and weak-to-moderate 
anisotropy regimes (for $r_x/r_y$ up to about 2,
which we verified by numerically solving the 2D 
Wannier-Schr\"{o}dinger equation), though its accuracy naturally 
drops in extreme anisotropic scenarios. 
Nevertheless, it reveals 
an important physical interplay between the effective mass and 
dielectric tensors: the binding energy scales logarithmically with 
the harmonic mean of the mass-polarizability products. 

Notice also that, in contrast to Keldysh's original 
asymptotic expansion (see discussion 
immediately following Eq. (8) in Ref. \cite{Keldysh1979}), our 
formulation provides an explicit, variationally-derived value 
of the constant energy shift, $ \ln 2 - \gamma_{\rm E} - 1$, that 
accompanies the dominant logarithmic contribution in 
Eq. (\ref{eq:Egs_Keldysh}).

\section{Conclusion}

We established a rigorous analytical framework for the strongly 
anisotropic Rytova-Keldysh potential by reducing its full 
two-dimensional inverse Fourier transform to an exact, non-linearized 
one-dimensional integral representation. Evaluating this exact form 
across all spatial regimes reveals a screening landscape that cannot 
be captured by simple isotropic averages. In the intermediate-range 
regime, the screening amplitude scales with $\sqrt{r_y(r_x - r_y)}$, 
demonstrating that the direction of weakest polarizability dominates 
the potential. In the short-range limit, $r \ll r_x, r_y$, the interaction 
reduces to an anisotropic logarithmic well governed by the geometric 
mean of the principal polarizabilities. This short-range collapse provides 
the formal mathematical justification for the numerical efficacy of 
the effective isotropic approximations noted in 
Ref.\ \cite{Kamban2020}. At macroscopic distances, 
$r \gg r_x, r_y$, the intrinsic two-dimensional anisotropy is completely 
screened out, restoring the familiar isotropic 3D Coulomb limit.

By exploiting the elliptical symmetry of the short-range logarithmic 
well, we constructed an anisotropic Gaussian variational ansatz 
to solve the 2D Wannier equation. The resulting closed-form 
expression for the exciton ground-state energy, Eq.\ (\ref{eq:Egs}), 
reveals the competing influences of the effective mass and 
dielectric tensors. The Gaussian trial function introduces very specific 
additive constant shift and recovers the standard $\ln(1/a_{\rm B})$ 
scaling behavior in the isotropic limit. Our analytical approach 
provides a reliable, computationally efficient approach, offering a direct 
predictive benchmark for investigating exciton stability and 
directional transport in low-symmetry 2D nanomaterials.

\appendix

\section{Fourier-Space Derivation of the Short-Range Asymptotics}
\label{sec:altVshort}

In the main text, the short-range logarithmic well was extracted 
from the exact 1D integral representation. For  
completeness, we provide an alternative derivation here by 
evaluating the large-momentum, $k \to \infty$, limit directly 
from the initial 2D Fourier representation.

At extremely short spatial distances, the local 2D screening 
completely dominates over the bulk dielectric background, 
corresponding to the condition $r_x k_x^2 + r_y k_y^2 \gg \kappa k$. 
In this regime, the $\kappa k$ term in the denominator of the 
full potential (\ref{eq:Vxy}) can be treated 
as a negligible perturbation, reducing the inverse Fourier transform 
to
\begin{equation}
V_{\rm short}(x, y) \approx \frac{qq'}{2\pi} 
\iint \frac{e^{i (k_x x + k_y y)}}{r_x k_x^2 + r_y k_y^2} dk_x dk_y.
\end{equation}
To evaluate this integral, we perform a momentum-space 
coordinate scaling to absorb the anisotropy. Defining the 
scaled momentum variables $q_x = k_x \sqrt{r_x}$ and 
$q_y = k_y \sqrt{r_y}$, the integral becomes the usual 
2D Coulomb potential in the scaled variables,
\begin{equation}
V_{\rm short}(x, y) \approx \frac{qq'}{2\pi \sqrt{r_x r_y}} 
\iint 
\frac{e^{i \left(q_x \frac{x}{\sqrt{r_x}} + q_y \frac{y}{\sqrt{r_y}}\right)}}{q_x^2 + q_y^2} 
dq_x dq_y.
\end{equation}
To evaluate this, we define an effective scaled radial coordinate 
$R = \sqrt{x^2/r_x + y^2/r_y}$ and transition to polar momentum 
coordinates $q_x = q \cos\varphi$ and $q_y = q \sin\varphi$. 
Because the pure $1/q^2$ integrand exhibits an infrared divergence 
at the origin, we must reintroduce a small momentum cutoff 
$q_c \sim \kappa/\sqrt{r_x r_y}$ to account for the macroscopic 
bulk screening term, $\kappa k$, that was dropped. 
The integration proceeds as follows:
\begin{align}
V_{\rm short}(x, y) 
&\approx 
\frac{qq'}{2\pi \sqrt{r_x r_y}} \int_0^{2\pi} d\varphi 
\int_{q_c}^\infty \frac{e^{i q R \cos(\varphi - \phi_R)}}{q^2} q \, dq 
\nonumber \\
&= 
\frac{qq'}{\sqrt{r_x r_y}} \int_{q_c}^\infty \frac{J_0(q R)}{q} dq 
= 
\frac{qq'}{\sqrt{r_x r_y}} \int_{q_c R}^\infty \frac{J_0(u)}{u} du 
\nonumber \\
&\approx 
\frac{qq'}{\sqrt{r_x r_y}} \left[ -\ln(q_c R) + C_0 \right] 
= 
-\frac{qq'}{2 \sqrt{r_x r_y}} \ln\left( \frac{x^2}{r_x} + \frac{y^2}{r_y} \right) + C,
\label{eq:Vshort_app}
\end{align}
where $J_0$ is the zeroth-order Bessel function of the first kind, 
$u = q R$ is a dimensionless integration variable, and all 
cutoff-dependent terms have been absorbed into the constant $C$. 

Equation (\ref{eq:Vshort_app}) serves as an independent 
Fourier-space confirmation of the main text results: at short 
ranges, equipotential contours are elliptical, and the effective 
dielectric screening length is governed by the geometric mean 
of the polarizabilities, $\sqrt{r_x r_y}$.

\section{Fourier-Space Derivation of the Long-Range Asymptotics}
\label{sec:altVlong}

Similarly, the long-range asymptotic behavior at macroscopic 
distances, $r \gg r_x, r_y$, can be independently verified by 
taking the small-momentum limit, $k \to 0$, of the 2D Fourier integral.

At large spatial separations, the electrostatic interaction extends 
far beyond the confines of the ultrathin 2D layer, meaning the 
potential must be dominated by the 3D dielectric environment. 
In momentum space, this implies the linear bulk dielectric screening 
term overwhelms the quadratic 2D polarizability tensor, such that 
$\kappa k \gg r_x k_x^2 + r_y k_y^2$. Under this condition, the inverse 
Fourier transform simplifies to
\begin{equation}
V_{\rm long}(x, y) \approx \frac{qq'}{2\pi} 
\iint \frac{e^{i (k_x x + k_y y)}}{\kappa \sqrt{k_x^2 + k_y^2}} dk_x dk_y.
\end{equation}
This is exactly the integral representation of the standard, unscreened 
Coulomb potential in a uniform dielectric medium. Correspondingly, 
integration gives
\begin{equation}
V_{\rm long}(r, \phi) \approx \frac{qq'}{\kappa \sqrt{x^2 + y^2}} 
= \frac{qq'}{\kappa r},
\end{equation}
showing that at macroscopic distances, the structural anisotropy of 
the 2D material completely washes out. The equipotential contours 
become circular, and the charges interact through a $1/r$ 
Coulomb potential mediated by the surrounding bulk dielectric 
constant $\kappa$.

\end{document}